\documentclass[10pt, final]{IEEEtran}

\usepackage{amsthm}

\usepackage{amssymb}
\usepackage{amsmath}
\usepackage{bbm}
\usepackage[ruled]{algorithm2e}
\usepackage{mathrsfs}
\usepackage{cite}
\usepackage{bm,epsfig,amsthm,url}
\usepackage{indentfirst}
\usepackage{amsfonts}
\usepackage{epstopdf}
\usepackage{cases}
\usepackage[caption=false,font=scriptsize]{subfig}
\usepackage{xcolor}
\usepackage{mathtools}

\makeatletter
\renewcommand{\maketag@@@}[1]{\hbox{\m@th\normalsize\normalfont#1}}%
\makeatother
\usepackage{hyperref}
\usepackage{stfloats}
\theoremstyle{definition}

\newtheorem{Proposition}{Proposition}

\begin{document}

\title{Hybrid Active-Passive IRS Assisted Energy-Efficient Wireless Communication}

\author{{Qiaoyan~Peng, Qingqing~Wu, Guangji~Chen, Ruiqi~Liu, Shaodan~Ma, Wen~Chen}
\thanks{Qiaoyan~Peng, Guangji~Chen and Shaodan~Ma are with the State Key Laboratory of Internet of Things for Smart City, University of Macau, Macao 999078, China (email: yc27464@umac.mo; guangjichen@um.edu.mo; shaodanma@um.edu.mo). Qingqing~Wu and Wen~Chen are with the
Department of Electronic Engineering, Shanghai Jiao Tong University, Shanghai 200240, China (e-mail: qingqingwu@sjtu.edu.cn; wenchen@sjtu.edu.cn). Ruiqi Liu is with the State Key Laboratory of Mobile Network and Mobile Multimedia Technology, ZTE Corporation, Shenzhen 518057, China (e-mail: richie.leo@zte.com.cn).}
}

\maketitle

\begin{abstract}
Deploying active reflecting elements at the intelligent reflecting surface (IRS) increases signal amplification capability but incurs higher power consumption. Therefore, it remains a challenging and open problem to determine the optimal number of active/passive elements for maximizing energy efficiency (EE). To answer this question, we consider a hybrid active-passive IRS (H-IRS) assisted wireless communication system, where the H-IRS consists of both active and passive reflecting elements. Specifically, we study the optimization of the number of active/passive elements at the H-IRS to maximize EE. To this end, we first derive the closed-form expression for a near-optimal solution under the line-of-sight (LoS) channel case and obtain its optimal solution under the Rayleigh fading channel case. Then, an efficient algorithm is employed to obtain a high-quality sub-optimal solution for EE maximization under the general Rician channel case. Simulation results demonstrate the effectiveness of the H-IRS for maximizing EE under different Rician factors and IRS locations.
\end{abstract}

\section{Introduction}
Intelligent reflective surface (IRS) has emerged as a revolution for future sixth-generation (6G) networks, which can be generally classified into two categories, i.e., the fully passive IRS and the fully active IRS \cite{Marco,RQLiu,chengzhi}. Specifically, the fully passive IRS can provide an asymptotic squared-power beamforming gain with low hardware cost \cite{jintao}, whereas it suffers from the ``multiplicative fading'' effect \cite{Dailinglong,DBLP:journals/twc/LongLPL21}. To overcome this issue, the fully active IRS has been proposed and investigated in \cite{DBLP:journals/twc/ChenWHCTJ23}. Since the fully active IRS requires refection-type amplifiers for each element, its total power consumption is much higher than its passive counterpart of the same size \cite{DBLP:journals/icl/ZhiPRCE22}. In summary, previous studies have shown that the conventional IRSs, i.e., the fully passive and active IRS, have complementary advantages.

A hybrid active-passive IRS (H-IRS) composed of both passive and active reflecting elements, has been recently proposed for further improving the performance beyond what can be achieved by using an active or passive IRS alone \cite{DBLP:conf/spawc/NguyenNWTCJ22}. The motivation for the need for H-IRS arises from the advantages and limitations of the conventional IRSs. As such, the \mbox{H-IRS} is a promising solution for enhancing various wireless systems, such as integrated sensing and communication (ISAC) systems \cite{ISAC} and unmanned aerial vehicle (UAV) communication \cite{HIRS_UAV}. In \cite{DBLP:conf/spawc/NguyenNWTCJ22}, the transmit precoder and \mbox{H-IRS} parameters were designed to maximize the sum rate of a multi-user system. The optimal elements allocation of the \mbox{H-IRS} for spectral efficiency (SE) maximization was explored in \cite{kang2022active} under the given deployment budget. In addition to SE, energy efficiency (EE) is also a major requirement of future 6G networks, which characterizes the fundamental trade-off between SE and system power consumption. Note that active elements introduce the new function of signal amplification, which is beneficial for improving SE, while they also require higher power consumption and hardware cost. As such, the operating region for the active IRS outperforms the passive IRS regarding EE, is not clear. Moreover, the H-IRS has the potential to balance the SE-cost trade-off by flexibly determining the number of active/passive elements. To this end, how to determine the number of active/passive elements for EE maximization is critical to make the H-IRS feasible for practical scenarios. 

\begin{figure}[t]
	\centering
	\includegraphics[width=3in]{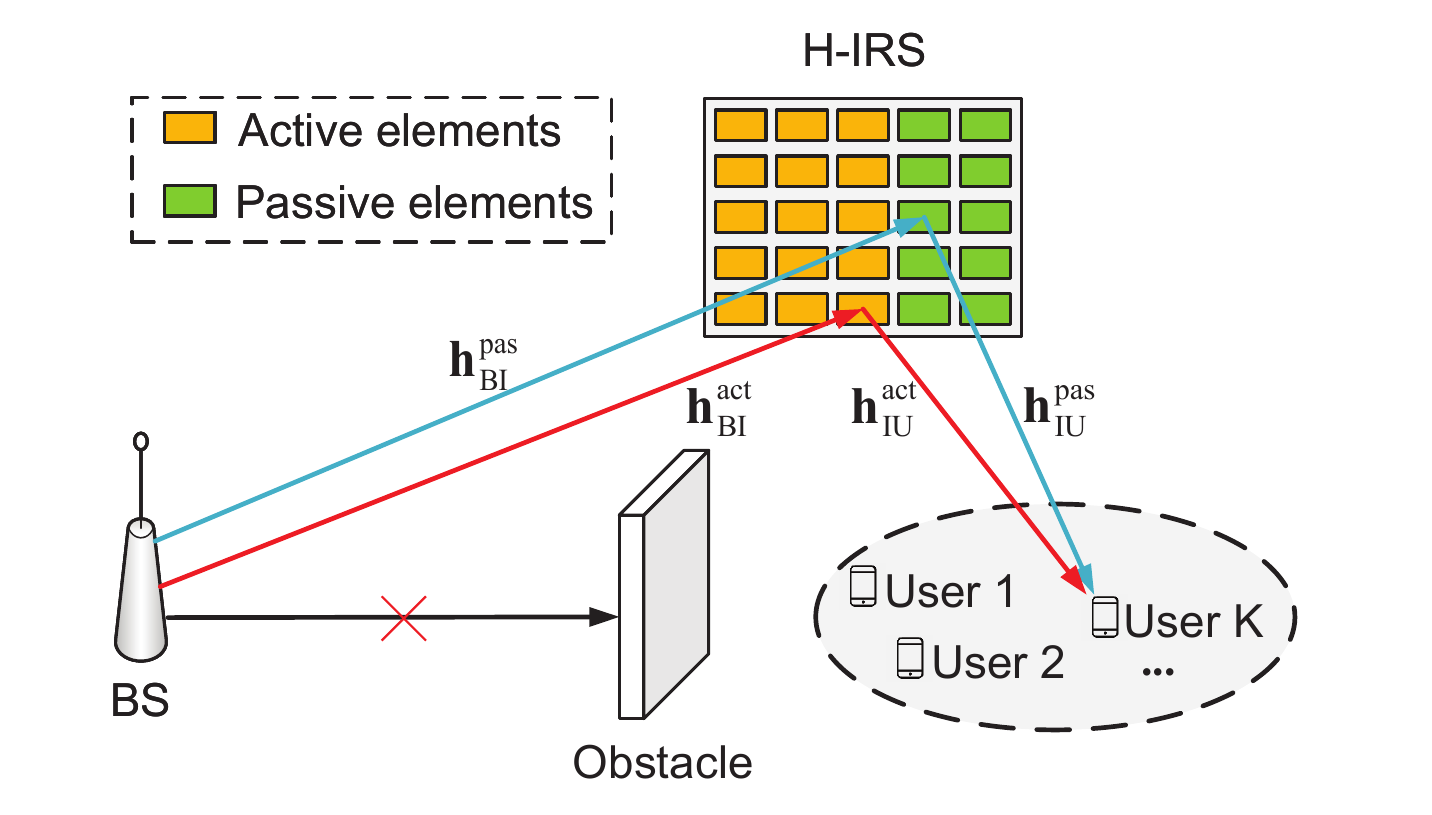}
        \vspace{-10pt}
	\caption{An H-IRS assisted wireless communication system.}
	\label{fig:systemmodel}
\vspace{-15pt}
\end{figure}

Motivated by the above considerations, we investigate the EE maximization problem in an H-IRS assisted wireless communication system. Specifically, we aim to determine the number of active/passive elements at the H-IRS to balance the trade-off between the power consumption incurred by active elements and the ergodic SE. 
The main contributions of this letter are summarized as follows: 
1) To obtain useful insights, we first consider two special cases, i.e., the line-of-sight (LoS) and Rayleigh fading channel cases. Under the LoS channel case, we derive the closed-form expression for a near-optimal solution. Furthermore, we show that at most one active element is required under the Rayleigh fading channel case.
2) We then propose an efficient algorithm to maximize the EE under the Rician fading channel case. 
3) Our numerical results demonstrate the EE of H-IRS with the optimized number of active/passive elements outperforms that of the fully active/passive IRSs under different Rician factors and IRS locations.

{\it{Notation:}} For a vector $\mathbf{x}$, $\left\| \mathbf{x} \right\|$, ${\left[ \mathbf{x} \right]}_{n}$ and $\arg ( \mathbf{x} )$ denote its Euclidean norm, $n$-th entry and phase vector, respectively. $\otimes$ denotes the Kronecker product. The distribution of a circularly symmetric complex Gaussian (CSCG) random variable with mean $\mu$ and variance $\sigma^2$ is denoted by $\mathcal{CN} ( \mu,\sigma^2 )$. $\lceil \cdot \rceil$ and $\lfloor \cdot \rfloor$ denote the ceiling and floor operations, respectively. $\arg \max ( \cdot )$ denotes the arguments at which the function value is maximized. 

\section{System Model and Problem Formulation}
\label{System Model and Problem Formulation}
As shown in Fig. \ref{fig:systemmodel}, we consider an H-IRS assisted wireless communication system composed of a single-antenna base station (BS), a cluster of single-antenna users and an H-IRS with ${N_{{\text{pas}}}}$ passive elements and ${N_{{\text{act}}}}$ active elements. To guarantee the overall system performance, we take the performance of the worst-case user into account.

We assume that the direct links between the BS and the users are blocked due to dense obstacles and IRS involved links follow the practical Rician fading model similar to \cite{DBLP:journals/icl/ZhiPRCE22,kang2022active}. The Rician fading model is able to capture the generalized channel environment with different LoS and non-LoS (NLoS) components by adjusting the Rician factor. As such, the equivalent baseband channel from the BS to the active IRS sub-surface is modeled as
${\mathbf{h}}_{{\text{BI}}}^{{\text{act}}} \!=\! \sqrt {{K_1}/({K_1} \!+\! 1)} {\bf{\bar h}}_{{\text{BI}}}^{{\text{act}}} \!+\! \sqrt {1/({K_1} \!+\! 1)} {\bf{\tilde h}}_{{\text{BI}}}^{{\text{act}}}$, where $K_1$ denotes the corresponding \mbox{Rician} fading factor. Specifically, the LoS component is expressed as ${\mathbf{\bar h}}_{{\text{BI}}}^{{\text{act}}} = {\beta_\text{BI}}{{\mathbf{a}}_r} ( {\theta _\text{BI}^r,\vartheta _\text{BI}^r,{N_{{\text{act}}}}} )$, where ${{\mathbf{a}}_r} (\! {\theta_{{\text{BI}}}^{\text{r}},\!\vartheta_{{\text{BI}}}^{\text{r}},\!{N_{{\text{act}}}}} \!) \!=\! {\mathbf{u}} ( {\frac{{2{d_{\text{I}}}}}{\lambda }\!\sin ( {\theta_{{\text{BI}}}^{\text{r}}} )\!\sin ({\vartheta_{{\text{BI}}}^{\text{r}}} )\!,\!{N_{x}}} )\! \otimes \!{\mathbf{u}} ( {\frac{{2{d_{\text{I}}}}}{\lambda }\!\cos \left( {\vartheta _{{\text{BI}}}^{\text{r}}} \right)\!,\!{N_{y}}} )$, $N \!=\! {N_x}{N_y}$, and ${\mathbf{u}} ( {\upsilon, \! M} ) = {[ {1,\ldots,{e^{ - ( {M - 1} )j\pi \upsilon }}} ]^T}$. ${{d_{\text{I}}}}$, $\lambda$ and $\beta_\text{BI}^2$ \mbox{denote} the element spacing of active elements, the wavelength and the path loss, respectively. $\theta _\text{BI}^r$ and $\vartheta _\text{BI}^r$ are the azimuth and \mbox{elevation} angles of arrival at the IRS, respectively. The NLoS \mbox{component} is given by ${[ {{\mathbf{\tilde h}}_{{\text{BI}}}^{{\text{act}}}} ] _n} \! \sim \!{\mathcal {CN}} ( {0,{\beta_\text{BI}^2}} ),\forall n \! \in \! {\mathcal N _{{\text{act}}}} \! \triangleq \left\{ {1, \cdots ,{N_{{\text{act}}}}} \right\}$. The equivalent baseband channel from the active IRS sub-surface to the worst-case user is modeled as
${\mathbf{h}}_{{\text{IU}}}^{{\text{act}}} \!=\! \sqrt {{K_2}/({K_2} \!+\! 1)} {\bf{\bar h}}_{{\text{IU}}}^{{\text{act}}} \!+\! \sqrt {1/({K_2} \!+\! 1)} {\bf{\tilde h}}_{{\text{IU}}}^{{\text{act}}}$ with the LoS component ${\mathbf{\bar h}}_{{\text{IU}}}^{{\text{act}}} \in {\mathbb{C}}^{{N_{{\text{act}}}} \times 1}$, the NLoS component ${\mathbf{\tilde h}}_{{\text{IU}}}^{{\text{act}}} \in {\mathbb{C}}^{{N_{{\text{act}}}} \times 1}$ and the corresponding Rician fading factor $K_2$. The equivalent baseband channel from the BS to the passive IRS sub-surface and from the passive IRS \mbox{sub-surface} to the worst-case user are denoted by ${{\mathbf{ h}}_{{\text{BI}}}^{{\text{pas}}}} \in {\mathbb{C}}^{{N_{{\text{pas}}}} \times 1}$ and ${{\mathbf{h}}_{{\text{IU}}}^{{\text{pas}}}} \in {\mathbb{C}}^{{N_{{\text{pas}}}} \times 1}$, which are defined similarly as ${{\mathbf{h}}_{{\text{BI}}}^{{\text{act}}}}$ and ${{\mathbf{h}}_{{\text{IU}}}^{{\text{act}}}}$.

Let ${{\mathbf{\Psi }}^{{\text{pas}}}} \triangleq {\operatorname{diag}} ( {{e^{j\varphi _1^{{\text{pas}}}}}, \cdots ,{e^{j\varphi _{{N_{{\text{pas}}}}}^{{\text{pas}}}}}} )$ denote the reflection matrix of the passive sub-surface, where $\varphi _n^{{\text{pas}}}$ represents the corresponding phase shift with $n \in {\mathcal{N}_{{\text{pas}}}} \buildrel \Delta \over = \{ {1, \cdots ,{N_{{\text{pas}}}}} \}$. The reflection matrix of the active sub-surface is denoted by ${{\mathbf{\Psi }}^{{\text{act}}}} \triangleq {{\mathbf{A}}^{{\text{act}}}}{{\mathbf{\Phi }}^{{\text{act}}}}$, where ${{\mathbf{A}}^{{\text{act}}}} \triangleq {\operatorname{diag}} \left( {{\alpha _1}, \cdots ,{\alpha _{{N_{{\text{act}}}}}}} \right)$ and ${{\mathbf{\Phi }}^{{\text{act}}}} \triangleq {\operatorname{diag}} ( {{e^{j\varphi _1^{{\text{act}}}}}, \cdots ,{e^{j\varphi _{{N_{{\text{act}}}}}^{{\text{act}}}}}} )$ denote its reflection amplification matrix and phase-shift matrix, respectively, with the amplification factor ${{\alpha _n}}$ and phase shift $\varphi _n^{{\text{act}}}$, $n \in {\mathcal{N_{{\text{act}}}}}$. Then, the \mbox{signal} received at the worst-case user is given by $y \!=\! {( {{\mathbf{h}}_{{\text{IU}}}^{{\text{act}}}} )^H}{{\mathbf{\Psi }}^{{\text{act}}}}{\mathbf{h}}_{{\text{BI}}}^{{\text{act}}}s \!+\! {( {{\mathbf{h}}_{{\text{IU}}}^{{\text{pas}}}} )^H}{{\mathbf{\Psi }}^{{\text{pas}}}}{\mathbf{h}}_{{\text{BI}}}^{{\text{pas}}}s + \! {( {{\mathbf{h}}_{{\text{IU}}}^{{\text{act}}}} )^H}{{\mathbf{\Psi }}^{{\text{act}}}}{\mathbf{n}_\text{r}} +\! {n_0}$, where $s \!\in \!\mathbb{C}$ denotes the transmitted data, which satisfies $\mathbb{E}\{ {{{\left| s \right|}^2}} \} = {P_\text{B}}$ with $P_\text{B}$ denoting the transmit power of the BS. $\mathbf{n}_\text{r} \! \sim \! {\mathcal {CN}} (\mathbf{0}_{N_{\text{act}}}, \sigma _\text{r}^2 \mathbf{I}_{N_{\text{act}}} )$ is the thermal noise introduced by active elements with the amplification noise power $\sigma _\text{r}^2$, and $n_0 \sim {\mathcal {CN}} \!(0 , \sigma _0^2 )$ is the additive white Gaussian noise (AWGN). Accordingly, the \mbox{signal-to-noise-ratio} (SNR) of the worst-case user is expressed as
\vspace{-5pt}
\begin{align}
\footnotesize
\gamma  = \frac{{{P_\text{B}}{{| {{{( {{\mathbf{h}}_{{\text{IU}}}^{{\text{act}}}} )}^H}{{\mathbf{\Psi }}^{{\text{act}}}}{\mathbf{h}}_{{\text{BI}}}^{{\text{act}}} + {{( {{\mathbf{h}}_{{\text{IU}}}^{{\text{pas}}}} )}^H}{{\mathbf{\Psi }}^{{\text{pas}}}}{\mathbf{h}}_{{\text{BI}}}^{{\text{pas}}}} |}^2}}}{{\sigma _\text{r}^2{{\| {{{( {{\mathbf{h}}_{{\text{IU}}}^{{\text{act}}}} )}^H}{{\mathbf{\Psi }}^{{\text{act}}}}} \|}^2} + \sigma _0^2}}.
\end{align}
As such, the ergodic achievable rate is given by
\begin{align}
R = \mathbb{E}\left\{ {{{\log }_2}\left( {1 + \gamma } \right)} \right\}. \label{R}
\end{align}

The average power consumption of the passive elements is given by ${P_{{\text{pas}}}} = {N_{{\text{pas}}}}{P_{\text{c}}}$ \cite{DBLP:journals/twc/LongLPL21}, where ${P_{\text{c}}}$ represents the switch and control circuit power consumption at each \mbox{reflecting} element. The average power consumption of the \mbox{active} \mbox{elements} is given by ${P_{{\text{act}}}} = {N_{{\text{act}}}}( {{P_{{\text{DC}}}} + {P_{\text{c}}}} ) + \xi \mathbb{E}\{ P_{{\text{out}}}\} ={N_{{\text{act}}}}( {{P_{{\text{DC}}}} + {P_{\text{c}}}} ) + \xi P_\text{I}$, where $\xi$ is the inverse of energy conversion coefficient and $P_{{\text{DC}}}$ is the DC biasing power consumption at each active element \cite{DBLP:journals/twc/LongLPL21}. ${P_{{\text{out}}}} = {P_\text{B}}{\left\| {{{\mathbf{\Psi }}^{{\text{act}}}}{\mathbf{h}}_{{\text{BI}}}^{{\text{act}}}} \right\|^2} + \sigma _\text{r}^2{\left\| {{{\mathbf{\Psi }}^{{\text{act}}}}} \right\|^2}$ denotes the output power of the active elements, which satisfies $\mathbb{E}\{ P_{{\text{out}}}\} = P_\text{I}$. As such, the average total power consumption is given by \cite{sub-connected}
\begin{align}
P_{\text{total}} = {P_{\text{pas}}} + {P_{\text{act}}} + {P_{{\text{BS}}}} + \varsigma {P_\text{B}}, \label{P_total}
\end{align}   
where $\varsigma$ and ${P_{{\text{BS}}}}$ are the inverse of energy conversion \mbox{coefficient} and the dissipated power consumed at the BS, respectively.  

We maximize the ergordic EE of the worst-case user by optimizing the number of active and passive elements, ${N_{{\text{act}}}}$ and ${N_{{\text{pas}}}}$, the IRS phase shifts, $\left\{ {{\mathbf{\Phi }}^{{\text{act}}}} \right.$, $\left. {{\mathbf{\Psi }}^{{\text{pas}}}} \right \}$, and the active-elements amplification matrix ${{\mathbf{A}}^{{\text{act}}}}$, which is defined as $\text{EE}  = {R}/{P_{\text{total}}}$. Given ${{\mathbf{A}}^{{\text{act}}}}$, ${N_{{\text{act}}}}$ and ${N_{{\text{pas}}}}$, i.e., the total power consumption is fixed, the EE maximization problem is equivalent to the SE maximization problem in \cite{kang2022active}. Accordingly, the optimal IRS phase shifts can be expressed as \cite{kang2022active}
\begin{align}
\varphi _n^{{\text{act}}} &= \arg \left( {{{ \left[ {{\mathbf{\bar h}}_{{\text{IU}}}^{{\text{act}}}} \right] }_n}} \right) - \arg \left( {{{\left[ {{\mathbf{\bar h}}_{{\text{BI}}}^{{\text{act}}}} \right] }_n}} \right),\forall n \in {\mathcal N_{{\text{act}}}},\\
\varphi _n^{{\text{pas}}} &= \arg \left( {{{\left[ {{\mathbf{\bar h}}_{{\text{IU}}}^{{\text{pas}}}} \right] }_n}} \right) - \arg \left( {{{\left[ {{\mathbf{\bar h}}_{{\text{BI}}}^{{\text{pas}}}} \right] }_n}} \right),\forall n \in {\mathcal N_{{\text{pas}}}}.
\end{align}
Similarly, given ${{\mathbf{\Phi }}^{{\text{act}}}}$, ${{\mathbf{\Psi }}^{{\text{pas}}}}$, ${N_{{\text{act}}}}$ and ${N_{{\text{pas}}}}$, the optimal amplification factor for the $n$-th active element is expressed as \cite{kang2022active}
\begin{align}
{\alpha _n} = {\alpha ^*} \triangleq \sqrt {{P_\text{I}}/\left({N_{{\text{act}}}{\left({{{P_\text{B}}{\beta_\text{BI}^2}} + \sigma _\text{r}^2}\right)}}\right) },\forall n \in {\mathcal N_{{\text{act}}}}.
\end{align}
Given ${{\mathbf{\Phi }}^{{\text{act}}}}$, ${{\mathbf{\Psi }}^{{\text{pas}}}}$, ${{\mathbf{A}}^{{\text{act}}}}$, the ergodic SE can be expressed as $\tilde{R} \!=\! {{{\log }_2}  ({1 \!+\!  \frac{{{{{P_\text{B}}{\beta_\text{BI}^2}{\beta_\text{IU}^2}}} ( {{\gamma _1}{{\left( {\sqrt {{A_{{\text{sum}}}}{N_{{\text{act}}}}} \!+\! {N_{{\text{pas}}}}} \right)}^2} \!+\! {\gamma _2} ( {{A_{{\text{sum}}}} \!+\! {N_{{\text{pas}}}}} )} )}}{{{{{A_{{\text{sum}}}}\sigma _\text{r}^2}}{\beta_\text{IU}^2} \!+\!  \sigma _0^2}}} )}$ \cite{kang2022active}. Then, the ergodic EE is given by
\begin{align}\label{EE}
&\eta ( {{N_{{\text{act}}}}, {N_{{\text{pas}}}}} ) \!=\! \tilde{R}/{P_\text{total}}\nonumber\\
=& \frac{{{{\log }_2}  (  {1 \!+\! \frac{{{{{P_\text{B}}{\beta_\text{BI}^2}{\beta_\text{IU}^2}}} ( {{\gamma _1}{{ ( {\sqrt {{A_{{\text{sum}}}}{N_{{\text{act}}}}} \!+\! {N_{{\text{pas}}}}} )}^2} \!+\! {\gamma _2} ( {{A_{{\text{sum}}}} \!+\! {N_{{\text{pas}}}}} )} )}}{{{{{A_{{\text{sum}}}}\sigma _\text{r}^2}}{\beta_\text{IU}^2} \!+\!  \sigma _0^2}}} )}}{{{N_{{\text{pas}}}}{P_{\text{c}}} \!+\! {N_{{\text{act}}}}  ( {{P_{{\text{DC}}}} \!+\! {P_{\text{c}}}} ) \!+\! \xi {{\mathcal{I}_{\mathbb{R}^{+} }}  ( {{N_{{\text{act}}}}} ) {P_{\text{I}}}} \!+\! {P_{{\text{BS}}}} \!+\! \varsigma  {P_\text{B}}}},
\end{align}
where ${A_{{\text{sum}}}} \triangleq {{{\mathcal{I}_{\mathbb{R}^{+} }}( {{N_{{\text{act}}}}} ){P_{\text{I}}}}}/{({{{{P_\text{B}}}}{\beta_\text{BI}^2} + \sigma _\text{r}^2})}$, and
\begin{align}
{\gamma _1} \triangleq \frac{{{K_1}{K_2}}}{{\left( {{K_1} + 1} \right)\left( {{K_2} + 1} \right)}},{\gamma _2} \triangleq \frac{{K_1}+{K_2}+1}{{\left( {{K_1} + 1} \right)\left( {{K_2} + 1} \right)}}. \label{y}
\end{align}
The indicator function ${\mathcal{I}_{\mathbb{R}^{+} }}\left( {{N_{{\text{act}}}}} \right) = 1$ if ${{N_{{\text{act}}}}} \textgreater 0$, otherwise ${\mathcal{I}_{\mathbb{R}^{+} }}\left( {{N_{{\text{act}}}}} \right) = 0$ .

Our objective is to maximize the ergodic EE of the \mbox{worst-case} user by optimizing the number of active/passive \mbox{elements}. Under the Rician fading channel case, the \mbox{optimization} problem is formulated as
\begin{align}
\label{P9}
\mathop {\max }\limits_{{N_{{\text{act}}}},{N_{{\text{pas}}}}}
\eta \left( {{N_{{\text{act}}}}, {N_{{\text{pas}}}}} \right)\;\;\;\;
{\text{s.t.}}\;\;{N_{{\text{act}}}} \in \mathbb{N},{N_{{\text{pas}}}} \in \mathbb{N}.
\end{align}
Problem \eqref{P9} is intractable because the discrete integer \mbox{variables} $\{{N_{{\text{act}}},N_{{\text{pas}}}}\}$ are coupled in the non-concave objective \mbox{function}. When ${N_{{\text{pas}}}}$ (${N_{{\text{act}}}}$) is zero, problem \eqref{P9} is reduced to the optimization problem with the fully active (passive) IRS. Thus, the proposed model generalizes the fully active and passive IRSs as two special cases.

\section{Proposed Solutions}
\label{Proposed Solutions}
In this section, we first investigate the active/passive \mbox{elements} allocation problem via two special cases, i.e., the LoS and Rayleigh fading channel cases, to draw important insights. In particular, the corresponding operating regions for active, passive, and hybrid IRS for EE maximization are characterized. Then, we propose an efficient algorithm to obtain its high-quality sub-optimal solution.

\subsection{LoS Channel Case} We first consider the LoS channel case with ${K_1} \to \infty$ and ${K_2} \to \infty$, which implies ${\gamma _1} \to 1$ and ${\gamma _2} \to 0$ from \eqref{y}. By relaxing the integer values ${N_{{\text{act}}}}$ and ${N_{{\text{pas}}}}$ into the continuous values ${x_{{\text{act}}}}$ and ${x_{{\text{pas}}}}$, the EE is given by
\begin{equation}
\small
{\eta _{{\text{L}}}}\!\left(\! {{x_{{\text{act}}}}, {x_{{\text{pas}}}}} \!\right) \!=\! \frac{{{{\log }_2}  (  {1 + \frac{{{{{P_\text{B}}{\beta_\text{BI}^2}{\beta_\text{IU}^2}}} {{{( {\sqrt {{A_{{\text{sum}}}}{x_{{\text{act}}}}} + {x_{{\text{pas}}}}} )}^2} } )}}{{{{{A_{{\text{sum}}}}\sigma _\text{r}^2}}{\beta_\text{IU}^2} +  \sigma _0^2})}} }}{{{x_{{\text{pas}}}} {P_{\text{c}}} \!\!+\!\! {x_{{\text{act}}}}  (\!{{P_{{\text{DC}}}} \!\!+\!\! {P_{\text{c}}}} \!) \!\!+\!\! \xi {{\mathcal{I}_{\mathbb{R}^{+}}}  \!(\! {{x_{{\text{act}}}}} \! )\!{P_{\text{I}}}} \!\!+\!\! {P_{{\text{BS}}}} \!\!+\!\! \varsigma  {P_\text{B}}}}.
\end{equation}
Then, problem \eqref{P9} is transformed to
\begin{align}
\label{EE_LoS_x}
\mathop {\max }\limits_{{x_{{\text{act}}}},{x_{{\text{pas}}}}} 
{\eta _{{\text{L}}}}\left( {{x_{{\text{act}}}},{x_{{\text{pas}}}}} \right)\;\;\;\;
{\text{s.t.}}\;\;{x_{{\text{act}}}} \in \mathbb{R}, {x_{{\text{pas}}}} \in \mathbb{R}.
\end{align}
We solve problem \eqref{EE_LoS_x} by considering three cases, namely the fully active, the fully passive, and the hybrid IRSs. By comparing the achievable EEs of the three cases, we obtain its near-optimal solution to problem \eqref{EE_LoS_x}.

\subsubsection{Fully Active IRS with ${x_{{\text{pas}}}} = 0$ and ${{x_{{\text{act}}}}} \textgreater 0$} We define ${\eta_\text{a} \left( {{x_{{\text{act}}}}} \right)} = {{{{\log }_2}( {1 + \beta _0}{x_{{\text{act}}}} )}}/{({{\beta _1}{x_{{\text{act}}}} + \beta _2})}$ and reformulate problem \eqref{EE_LoS_x} as
\begin{align}
\label{EE_act_x}
\mathop {\max }\limits_{{x_{{\text{act}}}}} {\eta_\text{a} \left( {{x_{{\text{act}}}}} \right)} \;\;\;\;
{\text{s.t.}}\;\;{x_{{\text{act}}}} \in \mathbb{R}^+,
\end{align}
where ${\beta _0} = {{{{{P_\text{B}}{\beta_\text{BI}^2}{\beta_\text{IU}^2}}}}/{({{{\sigma _\text{r}^2{\beta_\text{IU}^2}}} + {{\sigma _0^2}}/{{{A_{{\text{sum}}}^{'}}}}})}}$, ${\beta _1} = {P_{{\text{DC}}}} + {P_{\text{c}}}$, ${\beta _2} = \xi {P_{\text{I}}} + {P_{{\text{BS}}}} + \varsigma {P_\text{B}}\nonumber$ and ${A_{{\text{sum}}}^{'}} = {{{P_{\text{I}}}}}/{({{{{P_\text{B}}}}{\beta_\text{BI}^2} + \sigma _\text{r}^2})}$. 

\begin{Proposition}
\label{EE_act_trend}
${\eta_\text{a} (\! {{x_{{\text{act}}}}} \!)}$ first increases with ${{x_{{\text{act}}}}} \!\in\! (0,\!{x_{{\text{act}}}^*}]$ and then decreases with ${{x_{{\text{act}}}}} \!\in\! [{{x_{{\text{act}}}^*},\! \infty} \!)$. The optimal solution to problem \eqref{EE_act_x} is given by $x_{{\text{act}}}^* =\!  -  {1}/{{{\beta _0}}} ( {{{ (\! {{\beta _1} \!-\! {\beta _0}{\beta _2}} \!)}}/ (\!{{{\beta _1}L}} \!) \!+\! 1}  )$, where $L \!=\! \mathcal{W} ( {{{e^{ - 1}}( {{\beta _0}{\beta _2} - {\beta _1}} )}}/{{{\beta _1}}} )$ with the Lambert ${W}$ function $\mathcal{W} ( \cdot)$.
\end{Proposition}

{\it{Proof:}} The first-order derivative of ${\eta_\text{a} ( {{x_{{\text{act}}}}} )}$ with respect to (w.r.t.) $x_\text{act}$ is given by $\frac{{d ( {{\eta _{\text{a}}} ( {{x_{{\text{act}}}}}  )}  )}}{{d ( {{x_{{\text{act}}}}}  )}} \!=\! \frac{d_1 \left( {{x_{{\text{act}}}}} \right)}{{\ln 2{{ ( {{\beta _1}{x_{{\text{act}}}} + {\beta _2}} )}^2} ( {1 + {\beta _0}{x_{{\text{act}}}}} )}} \label{first-order}$, where
\begin{align}
\label{d1}
d_1 \!\left( \!{{x_{{\text{act}}}}}\! \right) \!=\! {{\beta _0}\!\left( \!{{\beta _1}{x_{{\text{act}}}} \!+\! {\beta _2}}\! \right) \!- \!\left(\! {1 \!+\! {\beta _0}{x_{{\text{act}}}}}\! \right)\!{\beta _1}\ln \! \left( \!{1 \!+\! {\beta _0}{x_{{\text{act}}}}} \!\right)}. 
\end{align}
Since $\frac{{{d}\left( {d_1 \left( {{x_{{\text{act}}}}} \right)} \right)}}{{d{{\left( {{x_{{\text{act}}}}} \right)}}}}  \textless 0$, $d_1 (\! {{x_{{\text{act}}}}} \!)$ monotonically decreases with $x_\text{act}$. There exists one and only one root ${x_\text{act}^\text{rt}}= - {1}/{{{\beta _0}}} ( {{{\left( {{\beta _1} - {\beta _0}{\beta _2}} \right)}}/{\left({{\beta _1}L}\right)} + 1} )$ for \eqref{d1}. When $0 \textless x_\text{act} \textless {x_\text{act}^\text{rt}}$, we have $d_1 ( {{x_{{\text{act}}}}} ) \textgreater 0$ and $\frac{{d\left( {\eta_\text{a} \left( {{x_{{\text{act}}}}} \right)} \right)}}{{d\left( {{x_{{\text{act}}}}} \right)}} \textgreater 0$, i.e., ${\eta_\text{a} ( {{x_{{\text{act}}}}} )}$ \mbox{monotonically} increases with $x_\text{act}$. When $x_\text{act} \textgreater {x_\text{act}^\text{rt}}$, we have $d_1 ( {{x_{{\text{act}}}}} ) \textless 0$ and $\frac{{d\left( {\eta_\text{a} \left( {{x_{{\text{act}}}}} \right)} \right)}}{{d\left( {{x_{{\text{act}}}}} \right)}} \textless 0$, i.e., ${\eta_\text{a} ( {{x_{{\text{act}}}}} )}$ monotonically decreases with $x_\text{act}$. Accordingly, ${\eta_\text{a} ( {{x_{{\text{act}}}}} )}$ is maximized at ${x_\text{act}^*} = {x_\text{act}^\text{rt}}$, which completes the proof. ~$\hfill\blacksquare$

\subsubsection{Fully Passive IRS with ${x_{{\text{act}}}} = 0$ and ${{x_{{\text{pas}}}}} \textgreater 0$} \mbox{Under} the practical scenarios, the number of IRS elements is very large, i.e., ${{x_{{\text{pas}}}}} \gg 1$. We define ${\eta_\text{p} \left( {{x_{{\text{pas}}}}} \right)} = {{{{\log }_2}\left( {{\beta _3} {x_{{\text{pas}}}}} \right)}}/{({{\beta _4}{x_{{\text{pas}}}} + {\beta _5}})}$ and \mbox{reformulate} \mbox{problem} \eqref{EE_LoS_x} as
\begin{align}
 \label{EE_pas_x}
\mathop {\max }\limits_{{x_{{\text{pas}}}}} {\eta_\text{p} \left( {{x_{{\text{pas}}}}} \right)}\;\;\;\;
{\text{s.t.}}\;\;{x_{{\text{pas}}}} \in \mathbb{R}^+,
\end{align}
where ${\beta _3} \!=\! \sqrt {{{{P_\text{B}}{\beta_\text{BI}^2}{\beta_\text{IU}^2}}}/{\sigma _0^2}}$, $
{\beta _4} \!=\! \frac{1}{2}{P_{\text{c}}}$ and $
{\beta _5} \!=\! \frac{1}{2}(\! {{P_{{\text{BS}}}} \!+\! \varsigma {P_\text{B}}}\!)$. Then, we have the following results.

\begin{Proposition}
\label{EE_pas_trend}
${\eta_\text{p} ( {{x_{{\text{pas}}}}} )}$ first increases with ${{x_{{\text{pas}}}}} \! \in \! (0,{x_{{\text{pas}}}^*}]$ and then decreases with ${{x_{{\text{pas}}}}} \in [{{x_{{\text{pas}}}^*}, \infty} )$. The optimal solution to problem \eqref{EE_pas_x} is given by $x_{{\text{pas}}}^* = {{{\beta _5}}}/{\left({{\beta _4}Q}\right)}$, where $Q \!=\! \omega (\! { -\! \ln ( {{{{\beta _4}}}/({{{\beta _3}{\beta _5}}}} )) \!-\! 1} )$ with the Wright omega function $\omega ( \cdot )$.
\end{Proposition}

{\it{Proof:}} The first-order derivative of ${\eta_\text{p} \left( {{x_{{\text{pas}}}}} \right)}$ w.r.t. $x_\text{pas}$ is given by $\frac{{d\left( {{\eta _{\text{p}}}\left( {{x_{{\text{pas}}}}} \right)} \right)}}{{d\left( {{x_{{\text{pas}}}}} \right)}} = \frac{d_2 \left( {{x_{{\text{pas}}}}} \right)}{\ln 2{\left( {{\beta _4}{x_{{\text{pas}}}} + {\beta _5}} \right)^2}{\beta _3}{{x_{{\text{pas}}}}}}$,
where
\begin{align}
d_2 \left( {{x_{{\text{pas}}}}} \right) = {\beta _3}\left({{{\beta _4}{x_{{\text{pas}}}} + {\beta _5}}}\right) - {\beta _4}{\beta _3}{x_{{\text{pas}}}}\ln \left( {{\beta _3}{x_{{\text{pas}}}}} \right). \label{d2}
\end{align}
Since $\frac{{{d}\left( {d_2 \left( {{x_{{\text{pas}}}}} \right)} \right)}}{{d{{\left( {{x_{{\text{pas}}}}} \right)}}}} \textless 0$, $d_2 ( {{x_{{\text{pas}}}}} )$ monotonically decreases with $x_\text{pas}$. There must exist one and only one root ${x_\text{pas}^\text{rt}} \!\!=\!\! {{{\beta _5}}}/{({{\beta _4}Q})}$ for \eqref{d2}. When $0 \textless x_\text{pas} \textless {x_\text{pas}^\text{rt}}$, we have $d_2 ( {{x_{{\text{pas}}}}} ) \textgreater 0$ and $\frac{{d( {\eta_\text{p} ( {{x_{{\text{pas}}}}} )} )}}{{d\left( {{x_{{\text{pas}}}}} \right)}} \textgreater 0$, i.e., ${\eta_\text{p} ( {{x_{{\text{pas}}}}} )}$ monotonically increases with $x_\text{pas}$. When $x_\text{pas} \textgreater {x_\text{pas}^\text{rt}}$, we have $d_2 ( {{x_{{\text{pas}}}}} ) \textless 0$ and $\frac{{d( {\eta_\text{p} ( {{x_{{\text{pas}}}}} )} )}}{{d( {{x_{{\text{pas}}}}} )}} \textless 0$, i.e., ${\eta_\text{p} ( {{x_{{\text{pas}}}}} )}$ monotonically decreases with $x_\text{pas}$. Accordingly, ${\eta_\text{p} ( {{x_{{\text{pas}}}}} )}$ is maximized at ${x_\text{pas}^*} = {x_\text{pas}^\text{rt}}$, which completes the proof. ~$\hfill\blacksquare$

\subsubsection{H-IRS with ${x_{{\text{act}}}} \textgreater 0$ and ${{x_{{\text{pas}}}}} \textgreater 0$}
We assume that $g_0^2{( {\sqrt {{A_{{\text{sum}}}^{'}}{N_{{\text{act}}}}} + {N_{{\text{pas}}}}} )^2} \gg 1$ and define ${\eta _{{\text{h}}}} ( {{x_{{\text{act}}}},{x_{{\text{pas}}}}} ) = {{2{{\log }_2}( {{g_0}}{( {\sqrt {{A_{{\text{sum}}}^{'}}{x_{{\text{act}}}}} + {x_{{\text{pas}}}}})} )}}/{({{x_{{\text{pas}}}}{P_{\text{c}}} + {\beta _1}{x_{{\text{act}}}} + {\beta _2}})}$, where ${g_0} = \sqrt {{{P_{\text{B}}}{\beta_\text{BI}^2}{\beta_\text{IU}^2}}/{{(\!{{{{A_{{\text{sum}}}^{'}}\sigma _\text{r}^2{\beta_\text{IU}^2}}} + \sigma _0^2}\!)}}}$. Then, problem \eqref{EE_LoS_x} is reformulated as
\begin{align}
\label{EE_hyb_x}
\mathop {\max } \limits_{{x_{{\text{act}}}},{x_{{\text{pas}}}}} 
{\eta _{{\text{h}}}}\left( {{x_{{\text{act}}}},{x_{{\text{pas}}}}} \right)\;\;\;\;
{\text{s.t.}}\;\;{x_{{\text{act}}}} \in \mathbb{R}^+,{x_{{\text{pas}}}} \in \mathbb{R}^+.
\end{align}

\begin{Proposition}
For any fixed ${x_{{\text{pas}}}}$, ${{\eta _{{\text{h}}}} ( {{x_{{\text{act}}}},{x_{{\text{pas}}}}} )}$ is a \mbox{quasi-concave} function w.r.t. ${x_{{\text{act}}}}$. Moreover, for any fixed ${x_{{\text{act}}}}$, ${{\eta _{{\text{h}}}} ( {{x_{{\text{act}}}},{x_{{\text{pas}}}}} )}$ is a quasi-concave function w.r.t. ${x_{{\text{pas}}}}$. As such, the optimal solution to problem \eqref{EE_hyb_x} is given by
\begin{equation}
{x_{{\text{h-a}}}^{{*}}} = {P_\text{c}^2}{A_{{\text{sum}}}^{'}}/{{\left( {2{\beta _1}} \right)}^2},\;\; 
{x_{{\text{h-p}}}^{{*}}} = \left( {{g_2} - \left( {1 + G} \right){g_1}} \right)/G, \label{x_hyb}
\end{equation}
where ${g_1} = {P_\text{c}}{A_{{\text{sum}}}^{'}}/( {2{\beta _1}} )$, ${g_2} = {g_1}/2 + {\beta_2}/{P_{\text{c}}}$ and $G = {\mathcal{W}}({{e^{ - 1}}( {{g_0}{g_2} - {g_0}{g_1}} )} )$.
\end{Proposition}

{\it{Proof:}} For any fixed ${x_{{\text{pas}}}}$, denote the upper contour set of ${{\eta _{{\text{h}}}}( {{x_{{\text{act}}}},{x_{{\text{pas}}}}} )}$ as ${S_{\alpha^{'}}} \!\!=\!\! \{ {{x_{{\text{act}}}} \!\in \!\mathbb{R}^{+} \vert {{\eta _{{\text{h}}}}( {{x_{{\text{act}}}},{x_{{\text{pas}}}}} )} \!\ge\! \alpha^{'}} \}$. ${S_{\alpha^{'}}}$ is equivalent to ${S_{\alpha^{'}}} \!\!=\!\! \{ {{x_{{\text{act}}}} \in \mathbb{R}^{+} \vert {\alpha^{'}} {U_{\alpha^{'}} }\!( {{x_{{\text{act}}}}} ) \!-\! {V_{\alpha^{'}} }\!( {{x_{{\text{act}}}}} ) \le 0 } \}$, where ${U_{\alpha^{'}} }\!( {{x_{{\text{act}}}}} )\!=\! {x_{{\text{pas}}}}{P_{\text{c}}} \!+\! {\beta _1}{x_{{\text{act}}}} \!+\! {\beta _2}$ and ${V_{\alpha^{'}} }\!( {{x_{{\text{act}}}}} )\!=\! 2{\log _2}({g_0}( {\sqrt {{A_{{\text{sum}}}^{'}}{x_{{\text{act}}}}}  \!+\! {x_{{\text{pas}}}}}))$. Since ${U_{\alpha^{'}} }\!( {{x_{{\text{act}}}}} )$ is linear and ${V_{\alpha^{'}} }\!( {{x_{{\text{act}}}}} )$ is concave, ${S_{\alpha^{'}}}$ is convex for any  $\alpha^{'} \in \mathbb{R}$. For any fixed ${x_{{\text{act}}}}$, denote the upper contour set of ${{\eta _{{\text{h}}}}( {{x_{{\text{act}}}},{x_{{\text{pas}}}}} )}$ as ${S_{\beta^{'}}} \!\!=\!\! \{ {{x_{{\text{pas}}}}\! \in \!\mathbb{R}^{+} \vert {{\eta _{{\text{h}}}}( {{x_{{\text{act}}}},{x_{{\text{pas}}}}} )} \ge \beta^{'}}\}$. Similarly, ${S_{\beta^{'}}}$ is \mbox{convex} for any $\beta^{'} \!\! \in \! \mathbb{R}$. ${{\eta _{{\text{h}}}}( {{x_{{\text{act}}}},{x_{{\text{pas}}}}} )}$ is quasi-concave if its \mbox{upper} \mbox{contour} set is convex. We set the partial derivative of ${{\eta _{{\text{h}}}}( {{x_{{\text{act}}}},{x_{{\text{pas}}}}})}$ w.r.t. ${x_{{\text{act}}}}$ and ${x_{{\text{pas}}}}$ to zero, i.e.,
\begin{align}
\small
\begin{cases}
\!\!\!{g_0}( {{P_{\text{c}}}{x_{{\text{pas}}}} + {\beta _1}{x_{{\text{act}}}} + {\beta _2}} ) \\
\!\!\!-\! {P_{\text{c}}}( {{g_0}\sqrt {{A_{{\text{sum}}}^{'}}{x_{{\text{act}}}}}  + {x_{{\text{pas}}}}} )\ln ( {{g_0}\sqrt {{A_{{\text{sum}}}^{'}}{x_{{\text{act}}}}}  + {x_{{\text{pas}}}}} ) \!=\! 0,\\
\!\!\!{g_0}\sqrt {{A_{{\text{sum}}}^{'}}} (\! {{P_{\text{c}}}{x_{{\text{pas}}}} + \!{\beta _1}{x_{{\text{act}}}} + {\beta _2}} \!) \\
\!\!\!- \! 2{\beta _1} \! \sqrt {{x_{{\text{act}}}}} ( {{g_0}\! \sqrt {{A_{{\text{sum}}}^{'}}{x_{{\text{act}}}}}  \!+\! {x_{{\text{pas}}}}} \!)\!\ln ( {{g_0}\! \sqrt {{A_{{\text{sum}}}^{'}}{x_{{\text{act}}}}}  \!+\! {x_{{\text{pas}}}}} \!) \!\!=\!\! 0. 
\end{cases} 
\end{align}
For any $x_\text{pas}$, the root $x_\text{act}^\text{rt}$ is unique. Given $x_\text{act}^\text{rt}$, there exists one and only one root $x_\text{pas}^\text{rt}$. The partial derivative of ${{\eta _{{\text{LoS}}}}(\! {{x_{{\text{act}}}^{\text{rt}}},{x_{{\text{pas}}}}}\!)}$ w.r.t. ${{x_{{\text{pas}}}}}$ is given by $\frac{{\partial ( {{\eta _{{\text{LoS}}}}( {{x_{{\text{act}}}^{\text{rt}}},{x_{{\text{pas}}}}} )} )}}{{\partial {x_{{\text{pas}}}}}} = \frac{{{d_3}( {{x_{{\text{pas}}}}} )}}{{2\ln 2( {{g_0}{g_1} + {g_0}{x_{{\text{pas}}}}} ){{( {{P_{\text{c}}}{x_{{\text{pas}}}} + {P_{\text{c}}}{g_1}/2 + {\beta _2}} )}^2}}}$, where
\begin{align}
\label{d3}
{d_3}( {{x_{{\text{pas}}}}} ) =& {g_0}( {{P_{\text{c}}}{x_{{\text{pas}}}} + {P_{\text{c}}}{g_1}/2 + {\beta _2}} ) \nonumber\\
&- {P_{\text{c}}}( {{g_0}{g_1} + {g_0}{x_{{\text{pas}}}}} )\ln ( {{g_0}{g_1} + {g_0}{x_{{\text{pas}}}}} ).
\end{align}
Since $\frac{{d( {{d_3}( {x_{{\text{pas}}}} )} )}}{{d( {x_{{\text{pas}}}} )}} \textless 0$, $d_3 ( {x_{{\text{pas}}}})$ monotonically decreases with ${x_{{\text{pas}}}}$. If ${g_0}( {{P_{\text{c}}}{g_1}/4 + {\beta _2}} ) - ( {{P_{\text{c}}}/2} )( {{g_0}{g_1}} )\ln ( {{g_0}{g_1}} ) \textless 0$, we have $d_3 ( {x_{{\text{pas}}}} ) \textless 0, \forall {x_{{\text{pas}}}}$, i.e., ${{\eta _{{\text{LoS}}}}( {{x_{{\text{act}}}^{\text{rt}}},{x_{{\text{pas}}}}})}$ monotonically \mbox{decreases} with ${x_{{\text{pas}}}}$. Accordingly, ${{\eta _{{\text{LoS}}}}( {{x_{{\text{act}}}^{\text{rt}}},{x_{{\text{pas}}}}})}$ is maximized at ${x_{{\text{pas}}}^*} = 0$. Otherwise, there must exist one and only one root ${x_{{\text{pas}}}^{\text{rt}}} = ( {{g_2} - ( {1 + G} ){g_1}} )/G$ for \eqref{d3}. When $0 \le {x_{{\text{pas}}}} \textless {x_{{\text{pas}}}^{\text{rt}}}$, we have $d_3 ( {x_{{\text{pas}}}} ) \textgreater 0$ and $\frac{{d( {{\eta _{{\text{LoS}}}}( {{x_{{\text{act}}}^{\text{rt}}},{x_{{\text{pas}}}}})} )}}{{d( {x_{{\text{pas}}}} )}} \textgreater 0$, i.e., ${{\eta _{{\text{LoS}}}}( {{x_{{\text{act}}}^{\text{rt}}},{x_{{\text{pas}}}}})}$ monotonically increases with ${x_{{\text{pas}}}}$. When ${x_{{\text{pas}}}} \textgreater {x_{{\text{pas}}}^{\text{rt}}}$, we have $d_3 ( {x_{{\text{pas}}}} ) \textless 0$ and $\frac{{d( {{\eta _{{\text{LoS}}}}( {{x_{{\text{act}}}^{\text{rt}}},{x_{{\text{pas}}}}})} )}}{{d( {x_{{\text{pas}}}} )}} \textless 0$, i.e., ${{\eta _{{\text{LoS}}}}( {{x_{{\text{act}}}^{\text{rt}}},{x_{{\text{pas}}}}})}$ monotonically decreases with $x_\text{pas}$. Accordingly, ${{\eta _{{\text{LoS}}}}( {{x_{{\text{act}}}^{\text{rt}}},{x_{{\text{pas}}}}})}$ is maximized at ${x_{{\text{pas}}}^*} = {x_{{\text{pas}}}^{\text{rt}}}$. In this case, ${\eta _{{\text{h}}}}( {{x_{{\text{act}}}},{x_{{\text{pas}}}}} )$ is maximized if and only if ${x_{{\text{act}}}}={x_\text{act}^\text{rt}}={x_{{\text{h-a}}}^{{*}}}$ and ${x_{{\text{pas}}}}=x_\text{pas}^\text{rt}={x_{{\text{h-p}}}^{{*}}}$, which are expressed in \eqref{x_hyb}. As such, the proof is completed. ~$\hfill\blacksquare$

Under the LoS channel case, the optimized number of active/passive elements and the architecture selection for the IRS (i.e., passive, active or hybrid) are given in \eqref{N_LoS} on the top of this page, where $\eta _{{\text{LoS}}}^* = \max ( {{\eta _{{\text{act}}}^*},{\eta _{{\text{pas}}}^*},{\eta _{{\text{hyb}}}^*} )}$. $\eta _{{\text{hyb}}}^*$\textgreater $\max ( {{\eta _{{\text{act}}}^*},{\eta _{{\text{pas}}}^*} )}$ is the operating region for the EE of H-IRS that outperforms that of the fully \mbox{active/passive} IRSs, where it is determined by the system parameters, i.e., the IRS location, the power \mbox{consumption}, etc. The maximum EE can be obtained at the IRS by flexibly determining the number of active/passive elements according to the system parameters. 

\begin{figure*}[ht]
	\begin{equation}
		\small
		\begin{cases}
			N_{{\text{pas}}}^* = 0,N_{{\text{act}}}^* = \mathop {\arg \max }\limits_{{x_{{\text{act}}}}} {\eta _{{\text{a}}}}\left( {{x_{{\text{act}}}}} \right), \text{if } \eta _{{\text{LoS}}}^* = {\eta _{{\text{act}}}^*} = \mathop { \max }\limits_{{x_{{\text{act}}}}} {\eta _{{\text{a}}}}\left( {{x_{{\text{act}}}}} \right),{x_{{\text{act}}}} \in \left\{ {\lfloor {x_{{\text{act}}}^*} \rfloor ,\lceil {x_{{\text{act}}}^*} \rceil } \right\}, \\
			N_{{\text{act}}}^* = 0,N_{{\text{pas}}}^* = \mathop {\arg \max }\limits_{{x_{{\text{pas}}}}} {\eta _{{\text{p}}}}\left( {{x_{{\text{pas}}}}} \right), \text{if } \eta _{{\text{LoS}}}^* = {\eta _{{\text{pas}}}^*} = \mathop { \max }\limits_{{x_{{\text{pas}}}}} {\eta _{{\text{p}}}}\left( {{x_{{\text{pas}}}}} \right),{x_{{\text{pas}}}} \in \left\{ {\lfloor {x_{{\text{pas}}}^*} \rfloor ,\lceil {x_{{\text{pas}}}^*} \rceil } \right\}, \\
			\! N_{{\text{act}}}^*,\! N_{{\text{pas}}}^* \!\!=\!\! \mathop {\arg \max }\limits_{{x_{{\text{act}}}},{x_{{\text{pas}}}}} {\eta _{{\text{h}}}}(\!  {{x_{{\text{act}}}},\!{x_{{\text{pas}}}}} \!) , \text{if } \eta _{{\text{LoS}}}^* \!\!=\!\! {\eta _{{\text{hyb}}}^*} \!\!=\!\!\! \mathop { \max }\limits_{{x_{{\text{act}}}},{x_{{\text{pas}}}}} \!\! {\eta _{{\text{h}}}}( \! {{x_{{\text{act}}}},\!{x_{{\text{pas}}}}} \! ),
			(\!  {{x_{{\text{act}}}},\! {x_{{\text{pas}}}}} \!) \! \in \! \{ \!{( {\lceil \!{x_{{\text{h-a}}}^{{*}}}\!\rceil ,\! \lceil\! {x_{{\text{h-p}}}^{{*}}} \!\rceil }  ) ,\! ( {\lceil\! {x_{{\text{h-a}}}^{{*}}}\! \rceil ,\! \lfloor\!  {x_{{\text{h-p}}}^{{*}}} \!\rfloor }  ) ,\! ( {\!\lfloor  {x_{{\text{h-a}}}^{{*}}} \!\rfloor ,\! \lceil\! {x_{{\text{h-p}}}^{{*}}} \!\rceil } ) ,\! (  {\lfloor\! {x_{{\text{h-a}}}^{{*}}} \!\rfloor ,\! \lfloor {x_{{\text{h-p}}}^{{*}}} \!\rfloor }  \!)} \! \} .
		\end{cases} \label{N_LoS}
	\end{equation}
	{\noindent} \rule[-0pt]{18.3cm}{0.05em}
\end{figure*}

\subsection{Rayleigh Fading Channel Case} \label{Sec_ray}
We next study the Rayleigh fading channel case with $K_1 = K_2 = 0$, which implies ${\gamma _1} = 0$ and ${\gamma _2} = 1$ from \eqref{y}. The EE under the Rayleigh fading channel case is given by
\begin{equation}
\footnotesize
\eta_\text{Ray} \!(\! {{N_{{\text{act}}}}, \!{N_{{\text{pas}}}}} \!) \!\!=\!\! \frac{{{{\log }_2}  (\!  {1 \!\!+\!\! {{{{{P_\text{B}}{\beta_\text{BI}^2}{\beta_\text{IU}^2}}} {(\! {{A_{{\text{sum}}}} \!\!+\!\! {N_{{\text{pas}}}}}\!) } }}\! / \!{(\!{{{{A_{{\text{sum}}}}\sigma _\text{r}^2}}{\beta_\text{IU}^2} \!\!+\!\!  \sigma _0^2}\!)}} \!)}}{{{N_{{\text{pas}}}}\!{P_{\text{c}}} \!\!+\!\! {N_{{\text{act}}}}  (\!{{P_{{\text{DC}}}} \!\!+\!\! {P_{\text{c}}}} \!) \!\!+\!\! \xi  {{\mathcal{I}_{\mathbb{R}^{+}}}  \!(\! {{N_{{\text{act}}}}} \! )\!{P_{\text{I}}}} \!\!+\!\! {P_{{\text{BS}}}} \!\!+\!\! \varsigma {P_\text{B}}}}. 
\end{equation}
Then, problem \eqref{P9} is reformulated as
\begin{align}
\label{EE_ray_N}
\mathop {\max }\limits_{{N_{{\text{act}}}},{N_{{\text{pas}}}}}{\eta_\text{Ray}\left( N_\text{act}, N_\text{pas} \right)} \;\;\;\;
{\text{s.t.}}\;\;{N_{{\text{act}}}} \in \mathbb{N},{N_{{\text{pas}}}} \in \mathbb{N}.
\end{align}
We solve problem \eqref{EE_ray_N} via two cases, i.e., $N_\text{act} = 0$ and $N_\text{act}  \textgreater 0$. First, we consider the case of $N_\text{act} \!=\! 0$ and define ${\eta _1}{({N_{{\text{p1}}}})} \!=\! {{{{\log }_2}(\! {1 \!\!+\!\! {{\beta _3^2}\!{N_\text{p1}}}} \!)}}\!/\!{({{N_\text{p1}}\!
{P_{\text{c}}} \!\!+\!\! {P_{{\text{BS}}}} \!\!+\!\! \varsigma {P_{\text{B}}}} \!)}$. Then, problem \eqref{EE_ray_N} is transformed to
\begin{align}
\label{EE_ray_N1}
\mathop {\max }\limits_{{N_\text{p1}}} {\eta _1}{\left({N_{{\text{p1}}}}\right)} \;\;\;\;{\text{s.t.}}\;\;{N_\text{p1}} \in \mathbb{N}.
\end{align}
By relaxing the integer value ${N _{{\text{p1}}}}$ into the continuous value ${x_{{\text{p1}}}}$, problem \eqref{EE_ray_N1} is converted to a convex problem. The optimal solution is given by ${x_{{\text{p1}}}^*} \!\!=\! \!-\! {1} \! /{{\beta _3^2}} \!(\! {{{( \!{{P_{\text{c}}} \!-\! 2\beta _3^2{\beta _5}} \!)}}\!/\!{(\!{{P_{\text{c}}}J}\!)} \!\!+\!\! 1}\!)$, where $J \!=\! {\mathcal{W}} (\! {{{{e^{ - 1}}( {2\beta _3^2{\beta _5} \!-\! {P_{\text{c}}}} )}}\!/\!{{{P_{\text{c}}}}}} \!)$. The optimal solution to problem \eqref{EE_ray_N1} is given by
\begin{align}
{N _{{\text{p1}}}^*} = \mathop {\arg \max }\limits_{x_{{\text{p1}}}} {\eta _1}{\left({x_{{\text{p1}}}}\right)}, x_{{\text{p1}}} \in \left\{ \lfloor {x_{{\text{p1}}}^*} \rfloor, \lceil {x_{{\text{p1}}}^*} \rceil \right\}.  
\end{align}

Second, for $N_\text{act} \textgreater 0$, we obtain that $N_\text{act}^{*} = 1$ because ${\eta_\text{Ray}(\! N_\text{act}, N_\text{pas} \!)}$ monotonically decreases with $N_\text{act}$. We define ${{\eta _2}( {{N_{{\text{p2}}}}} )} \!=\! {{{{\log }_2} ( {1 \!+\! {{{\beta _0}} ( {1 \!+\! {A_{{\text{sum}}}^{'}}{N _{{\text{p2}}}}} )}} )}}/{({{N _{{\text{p2}}}}{P_{\text{c}}} \!+\! {\beta _6}})}$ and reformulate problem \eqref{EE_ray_N} as
\begin{align}
\label{EE_ray_N2}
\mathop {\max }\limits_{{N _{{\text{p2}}}}}  
{{\eta _2}( {{N_{{\text{p2}}}}} )}\;\;\;\;{\text{s.t.}}\;\;{N _{{\text{p2}}}} \in \mathbb{N},
\end{align}
where $\beta _6 = \beta _1 + \beta _2$. By relaxing the integer value $N _{{\text{p2}}}$ into the continuous value $x_{{\text{p2}}}$, problem \eqref{EE_ray_N2} is converted to a \mbox{convex} problem. The optimal solution is given by ${x_{{\text{p2}}}^{*}} = 0$ if ${{\beta _0}{\beta _6}}/{{{A_{{\text{sum}}}^{'}}}} - {P_{\text{c}}}( {1 + {\beta _0}} )\ln ({1 + {\beta _0}})$\textless 0; otherwise, ${x_{{\text{p2}}}^{*}} = {A_{{\text{sum}}}^{'}}/{\beta _0} (({\beta _0}{\beta _6}/{A_{{\text{sum}}}^{'}} \!-\! (1 \!+\! {\beta _0}){P_\text{c}})/({P_\text{c}}{L^{'}}) \!-\! (1 \!+\! {\beta _0}) )$, where ${L^{'}} \!=\!\mathcal{W} (\! {{e^{-1}}( {{\beta _0}{\beta _6} / {A_{{\text{sum}}}^{'}} \!-\! ( {1 \!+\! {\beta _0}} ){P_{\text{c}}}})/{P_{\text{c}}}} \!)$. The optimal solution to problem \eqref{EE_ray_N2} is given by
\begin{align}
{N _{{\text{p2}}}^*} = \mathop {\arg \max }\limits_{x_{{\text{p2}}}} {{\eta _2}\left( {{x_{{\text{p2}}}}} \right)}, {{x_{{\text{p2}}}}} \in \left\{ \lfloor {{x_{{\text{p2}}}^*}} \rfloor, \lceil {{x_{{\text{p2}}}^*}} \rceil \right\}.
\end{align}

Based on the previous discussions, the optimal solution to problem \eqref{EE_ray_N} is expressed as
\begin{align}
\label{rayleigh_proposition}
\left\{ {\begin{array}{*{20}{l}}
{{N_{{\text{act}}}^*} = 0,{N_{{\text{pas}}}^*} = {N _{{\text{p1}}}^*},}\\
{{N_{{\text{act}}}^*} = 1,{N_{{\text{pas}}}^*} = {N _{{\text{p2}}}^*},}
\end{array}\begin{array}{*{20}{l}}
{\eta _\text{p1}^{*}} \ge {\eta _\text{p2}^{*}},\\
\text{Otherwise,}
\end{array}} \right. 
\end{align}
where ${\eta _\text{p1}^{*}} = {\eta _1}{({N_{{\text{p1}}}^{*}})}$ and ${\eta _\text{p2}^{*}} = {\eta _2}{({N_{{\text{p2}}}^{*}})}$. It is observed from \eqref{rayleigh_proposition} that at most one active element is required under the Rayleigh fading channel case. In this case, the system cannot attain beamforming gain since the design of IRS phase shifts is based only on the LoS components. Moreover, the active elements also cannot reap aperture gain due to the amplified power constraint. Therefore, the ergodic rate is independent of $N_\text{act}$ and thus deploying more active elements only results in higher power consumption rather than the improvement of rate, which leads to the fact that at most one active element is needed. By contrast, the ergodic rate still scales linearly w.r.t. $N_\text{pas}$ benefited from the aperture gain provided by passive elements.

\vspace{-5pt}
\subsection{Rician Fading Channel Case}
Finally, we study the general Rician fading channel case. Since the discrete integer variables $\{{N_{{\text{act}}}},{N_{{\text{pas}}}}\}$ are \mbox{coupled} in the objective function, problem \eqref{P9} is a \mbox{non-convex} \mbox{optimization} problem, which is challenging to be solved optimally. To overcome this issue, we obtain the following proposition.

\begin{Proposition}
\label{EE_Rician_prop}
For any fixed ${x_{{\text{pas}}}}$, ${{\eta}\left( {{x_{{\text{act}}}},{x_{{\text{pas}}}}} \right)}$ is a \mbox{quasi-concave} function w.r.t. ${x_{{\text{act}}}} \in \mathbb{R}^{+}$, where $x_\text{act}$ and $x_\text{pas}$ are the continuous values of $N_\text{act}$ and $N_\text{pas}$ with integer relaxation, respectively. 
\end{Proposition}

{\it{Proof:}} For any fixed ${x_{{\text{pas}}}}$, denote the upper contour set of ${{\eta} ( {{x_{{\text{act}}}},{x_{{\text{pas}}}}} )}$ as ${S_{\tau}} \!\!=\!\! \{ {{x_{{\text{act}}}} \in \mathbb{R}^{+} \vert {{\eta}( {{x_{{\text{act}}}},{x_{{\text{pas}}}}} )} \ge \tau} \}$. ${S_{\tau}}$ is equivalent to ${S_{\tau}} \!\!=\!\! \{ {{x_{{\text{act}}}} \in \mathbb{R}^{+} \vert \tau {U_\tau }( {{x_{{\text{act}}}}} ) \!-\! {V_\tau }( {{x_{{\text{act}}}}} ) \le 0 } \}$, where ${V_\tau }( {{x_{{\text{act}}}}} )\!=\! {\log _2}(1 \!+\! {P_{\text{B}}}\beta _{{\text{BI}}}^2\beta _{{\text{IU}}}^2({\gamma _1}{(\sqrt {{A_{{\text{sum}}}}{x_{{\text{act}}}}}  \!+\! {x_{{\text{pas}}}})^2} \!+\! {\gamma _2}({A_{{\text{sum}}}} \!+\! {x_{{\text{pas}}}}))/( {{A_{{\text{sum}}}}\sigma _{\text{r}}^2\beta _{{\text{IU}}}^2 \!+\! \sigma _0^2} ))$ and ${U_\tau }( {{x_{{\text{act}}}}} )\!=\! {{x_{{\text{pas}}}}{P_{\text{c}}} \!+\! {x_{{\text{act}}}}{\beta_1} \!+\! \xi {P_{\text{I}}} \!+\! 2{\beta_5}}$. Since ${V_\tau }\!( {{x_{{\text{act}}}}} )$ is concave and ${U_\tau }\!( {{x_{{\text{act}}}}} )$ is linear, ${S_{\tau}}$ is convex for any  $\tau  \! \in \! \mathbb{R}$. As such, the proof is completed. ~$\hfill\blacksquare$

Based on Proposition \ref{EE_Rician_prop}, we can apply an efficient algorithm to obtain a high-quality sub-optimal solution under the Rician fading channel case. Given ${x_{{\text{pas}}}}$, we can obtain the optimal solution ${x_{{\text{act}}}^* \!\in\! \mathbb{R}^{+}}$ by the Newton's algorithm or ${x_{{\text{act}}}^*}\!=\!0$. Given ${x_{{\text{act}}}^*}$, one stationary point of ${x_{{\text{pas}}}}$ can be obtained by adopting the gradient ascent method. By updating ${x_{{\text{pas}}}}$ and ${x_{{\text{act}}}}$ iteratively until the convergence is reached, one high-quality solution to the original problem can be obtained.

\vspace{-5pt}
\begin{figure*}[ht]
        \centering
	\begin{minipage}[t]{0.49\linewidth}
        \subfloat[EE versus $K$.]
        {\label{fig:K_EE}\includegraphics[width=0.5\textwidth]{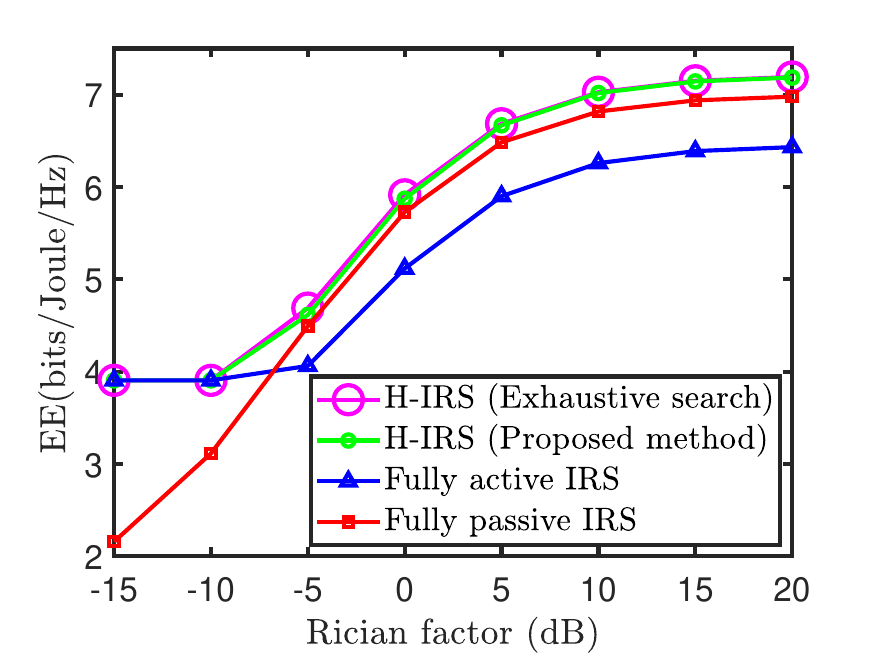}}
        \hspace{-0.21in}
        \subfloat[Optimized $N_\text{act}$ and $N_\text{pas}$ versus $K$.]{\label{fig:K_Nact_Npas}\includegraphics[width=0.5\textwidth]{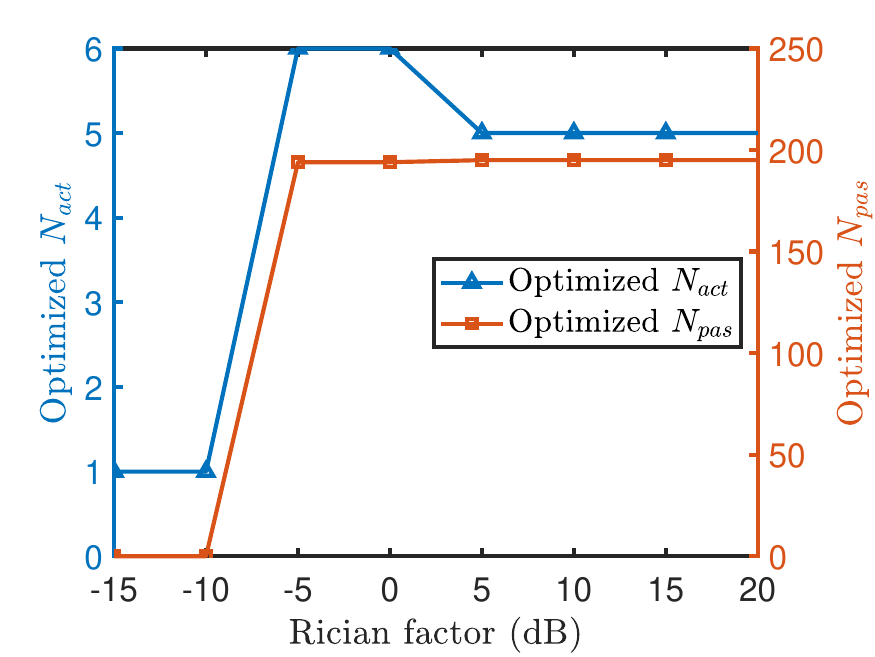}}
        \caption{Comparison on EE and elements allocation of the H-IRS versus $K$.}
		\label{K_EE}
	\end{minipage}
        \hspace{0.05in}
	\begin{minipage}[t]{0.49\linewidth}
        \subfloat[EE versus $x_\text{IRS}$.]{\label{fig:d_EE}\includegraphics[width=0.5\textwidth]{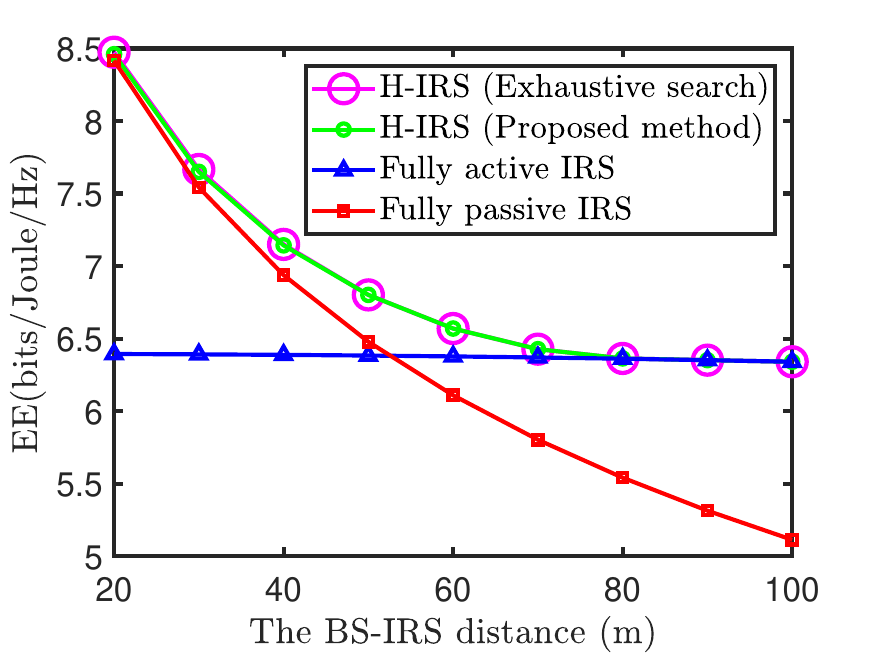}}
        \hspace{-0.15in}
        \subfloat[Optimized $N_\text{act}$ and $N_\text{pas}$ versus $x_\text{IRS}$.]{\label{fig:d_Nact_Npas}\includegraphics[width=0.5\textwidth]{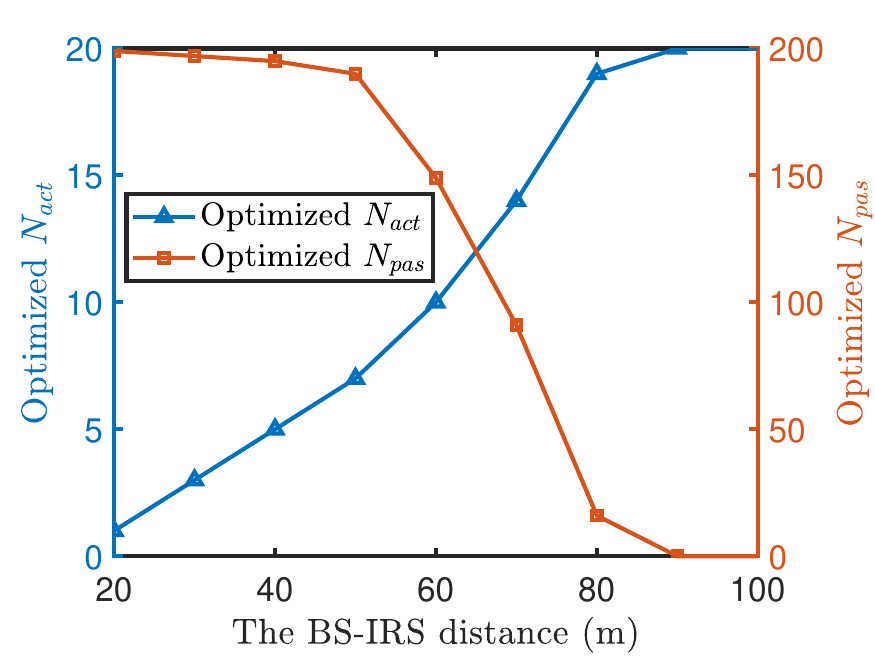}}
        \caption{Comparison on EE and elements allocation of the H-IRS versus $x_\text{IRS}$.}
		\label{d_EE}
	\end{minipage}
\end{figure*}

\vspace{10pt}
\section{Simulation results}
In this section, numerical results are provided to illustrate the effectiveness of using the H-IRS to improve the \mbox{ergodic} EE. The BS, the H-IRS and the worst-case user are located at $(0,0)$ meter (m), $(x_\text{IRS}, 0)$ m, and $(x_\text{IRS}, 10)$ m, respectively. For both the BS-IRS and IRS-user links, the Rician factors are considered to be the same, i.e., $K_1 = K_2 = K$, the path-loss factors are set to 2.2 and the signal attenuation at a reference distance of \mbox{1 m} is set to 30 dB. Other system parameters are set as follows: $\sigma _\text{r}^{2}\!=\!\sigma _0^{2}\!=\! -80$ dBm, $P_\text{B}\!=\!20$ dBm, $P_\text{I}\!=\!-15$ dBm, $P_\text{C}=1.5$ dBm, $P_\text{DC}=10$ dBm, $P_\text{BS}=30$ dBm, and $\xi=\varsigma=1.1$.

In Fig. \ref{fig:K_EE}, we plot the ergodic EE of the worst-case user versus Rician factor when $x_\text{IRS}=40$ m. It is observed that the EE of H-IRS with the active/passive elements optimized by the proposed method is close to that by the exhaustive search algorithm, which outperforms that of the fully active/passive IRS under different Rician factors. The reason is that the \mbox{H-IRS} provides an additional degree of freedom for determining the number of active/passive elements, thereby enabling a flexible balance the trade-off between SE and power consumption. In Fig. \ref{fig:K_Nact_Npas}, we plot the number of passive and active elements of the H-IRS optimized by the proposed algorithm versus Rician factor $K$. One can observe that the optimized $N_\text{act}$ and $N_\text{pas}$ are 5 and 195, respectively when $K = 15$ dB, while the optimized $N_\text{act}$ is one and the optimized $N_\text{pas}$ is zero when $K \le 10$ dB, which agrees with our analysis in Section \ref{Sec_ray}. Then, the EE of the \mbox{H-IRS} is equal to that of the fully active IRS (see Fig. \ref{fig:K_EE}). In addition, it is observed that the required number of active elements first increases and then decreases with the Rician factor. This is because the received power scales proportionally w.r.t. $\gamma_1 N_\text{pas}^2$ and thus increases significantly with $N_\text{pas}$ when $K$ is not very small, i.e., $\gamma_1$ approaches 1. Benefiting from the high passive beamforming gain in the LoS-dominated channel case, the system tends to employ fewer active elements to reduce power consumption, which is helpful for maximizing EE.

In Fig. \ref{fig:d_EE}, we plot the ergodic EE of the worst-case user versus $x_\text{IRS}$ when $K=15$ dB. First, it is observed that the EE of the three IRSs decrease with $x_\text{IRS}$. Second, we observe that the EE of the fully passive IRS decreases significantly and that of the fully active IRS remains almost unchanged with $x_\text{IRS}$. The reason is that the fully passive IRS suffers from severe path loss attenuation while the fully active IRS can amplify the signal attenuated after the transmission via the BS-IRS link. In Fig. \ref{fig:d_Nact_Npas}, we plot the number of passive and active elements of the H-IRS optimized by the proposed algorithm versus the IRS location $x_\text{IRS}$. It is observed that the optimized $N_\text{act}$ is 7 and the optimized $N_\text{pas}$ is 190 when $x_\text{IRS} = 50$ m, while the optimized $N_\text{act}$ and $N_\text{pas}$ are 20 and 0, respectively when \mbox{$x_\text{IRS} \ge 90$ m}. Then, the EE of the H-IRS is equal to that of the fully active IRS (see Fig. \ref{fig:d_EE}). In addition, it is observed that the optimized number of active elements increases and that of passive elements decreases with $x_\text{IRS}$. It is because more active elements should be deployed when the system suffers severe path loss attenuation, thereby improving the EE.

\section{Conclusion}
\label{Conclusion}
This letter studied the elements allocation problem for \mbox{maximizing} the ergodic EE in an H-IRS assisted \mbox{wireless} \mbox{communication} system. We first derived the \mbox{closed-form} \mbox{expression} for a near-optimal solution under the LoS \mbox{channel} case and unveiled that at most one active element is \mbox{required} under the Rayleigh channel case. Then, we proposed an \mbox{efficient} algorithm under the general Rician fading \mbox{channel} case. \mbox{Simulation} results demonstrated that the H-IRS is a promising architecture for flexibly balancing the SE-cost \mbox{trade-off}. In future work, it is worthy of further investigating the EE of \mbox{H-IRS} assisted optical wireless networks.

\bibliographystyle{IEEEtran}
\bibliography{refs.bbl}

\end{document}